# Various Architectures of Colloidal $Cu_3(MoO_4)_2(OH)_2$ and $Cu_3Mo_2O_9$; Thermal Stability, Photoluminescence and Magnetic Properties of $Cu_3(MoO_4)_2(OH)_2$ and $Cu_3Mo_2O_9$ Nanosheets

Azam Bayat [a,*], Ali Reza Mahjoub [a], Mostafa M. Amini [b]

[a]*Department of Chemistry, Tarbiat Modares University, Tehran 14155-4383, Iran*

[b]*Department of Chemistry, Shahid Beheshti University, G. C., Tehran 198396311, Iran*

**Corresponding author: Azam Bayat**

**E-mail address: azam.bayat @modares.ac.ir**

**Abstract**

The lindgrenite compounds [$Cu_3(MoO_4)_2(OH)_2$] with various architectures and high crystallinity were prepared by a simple surfactant-assisted hydrothermal method. Then, the $Cu_3Mo_2O_9$ samples were prepared by calcination of the as-synthesized $Cu_3(MoO_4)_2(OH)_2$. The resulting samples have high crystallinity, colloidal properties, high-yield, large-scale production capability with using of nontoxic and inexpensive reagents and water as an environmentally solvent. The scanning electron microscope studies show that the as-prepared lindgrenite nanostructures are well crystallized with rod, sheet and hollow sphere morphologies. These products were content of the $Cu_3(MoO_4)_2(OH)_2$ rods with diameters of about 100 nm, the $Cu_3(MoO_4)_2(OH)_2$ nanosheets with thickness of 30–100 nm and the $Cu_3(MoO_4)_2(OH)_2$ hallow spheres, consisting of a large number of nanosheets with thickness of about 40-70 nm. The $Cu_3Mo_2O_9$ samples that obtained from thermal treatment of lindgrenite retained the original morphologies. Meanwhile, the photoluminescence and magnetic properties of the nanosheet samples have been investigated that the both of $Cu_3(MoO_4)_2(OH)_2$ and $Cu_3Mo_2O_9$ samples have super paramagnetic behavior at room temperature and in comparison with previous works, $Cu_3(MoO_4)_2(OH)_2$ and $Cu_3Mo_2O_9$ samples synthesized by the surfactant-assisted hydrothermal method in this work have a very obvious red-shifted PL emission and high intensity.

*Keywords*: $Cu_3(MoO_4)_2(OH)_2$; $Cu_3Mo_2O_9$; colloidal; super paramagnetic; red emission

**Introduction**

Molybdenum oxide-based materials, especially binary ones, have been received considerable attention due to their applications in catalysts, display devices, photochromisms, sensors, batteries, absorption, electrical conductivity, magnetism, photochemistry and smart windows [1-2].

Lindgrenite is one of the rare minerals that originally found in Chile resulting from the oxidation of primary molybdenite [3]. Preparation of natural minerals is useful for verifying the quality of minerals and detection the geological origin of mineral formation. The structural and composition determination of the synthetic mineral crystals may establish the standards for evaluating the quality of the natural minerals. Furthermore, the synthetic approach may facilitate the large-scale production of low-cost, high-quality natural minerals, in particular, those that have technological and economic interest but are rare or impure in nature. Study of the conditions for mineral formation helps mineralogists to explore the minerals in specific zones, where the geological reactions once occurred under the similar circumstance mentioned above [4]. Lindgrenite has been shown to be an effective fire retardant and smoke suppressor when combined with $CuSnO_3$ in polyvinyl chloride plastic (PVC). For this purpose, it seems that the needle-like and the sheet-like structure materials are good candidates for functional polymeric composites and fiber hybrid materials [5-7].

On the other hand, the properties of $Cu_3Mo_2O_9$ have attracted extensive interests in various research fields [1]. Molybdenum trioxide nanostructures can be used as catalysts. For instance, $Cu_3Mo_2O_9$ is employed in the deep oxidation of diesel exhaust soot with high activity, but reports on $Cu_3Mo_2O_9$ morphologies are very few [2, 8].

Shores et al. have reported the synthesis of lindgrenite elegantly modified by bipyridine and piperazine. The two ditopic ligands, interestingly, break the inorganic 3D network into layers, consequently transforming the ferromagnetic lindgrenite into antiferromagnets. Vilminot et al. presented the hydrothermal synthesis of the hydrogenated and deuterated analogues of lindgrenite [9]. Bao et al. synthesized $Cu_3(MoO_4)_2(OH)_2$ which could be transferred to metastable $Cu_3Mo_2O_9$ by thermal decomposition. Xu et al. prepared $Cu_3Mo_2O_9$ with hollow and prickly sphere-like architecture by a simple thermal treatment of lindgrenite. Also, Jiang et al. synthesized lindgrenite nanocrystals by simple aqueous precipitation, which could be decomposed to $Cu_3Mo_2O_9$. Hasan et al. synthesized $Cu_3Mo_2O_9$ for catalytic oxidation of the harmful gases and Zhang et al. prepared $Cu_3Mo_2O_9$ nanosheets by a hydrothermal method and investigate the direct electrochemistry of hemoglobin (Hb). Finally, Xia et al. constructed $Cu_3Mo_2O_9$ nanoplates with excellent lithium storage properties based on a pH-dependent dimensional change [1, 10].

In this work, by a hydrothermal surfactant-assisted method, the colloidal $Cu_3(MoO_4)_2(OH)_2$ and $Cu_3Mo_2O_9$ nanorods, nanosheets and self-assembled hallow spheres with high crystallinity were synthesized. A significant feature of this synthetic approach was pointed out that the resulting samples have high crystallinity, hydrophilic surface, high-yield, large-scale production capability with using of nontoxic and inexpensive reagents and water as an environmentally benign reaction solvent. For use these nanostructures in applied sciences it is must to disperse or soluble in water and the solubility/dispersibility of the samples plays a key role in their applications. The size and shape of the as-synthesized samples can be controlled readily by simple tuning the reaction parameters and the reaction parameters play a major role in the morphologies of the final samples. Meanwhile, possible growth mechanisms for the formation of single-crystalline $Cu_3(MoO_4)_2(OH)_2$ products were proposed. Moreover, the photoluminescence and magnetic properties of the as-synthesized $Cu_3(MoO_4)_2(OH)_2$ and $Cu_3Mo_2O_9$ nanosheets were determined in which a red emission was observed at 706 nm by excitation at 237 nm in their photoluminescence spectra of the $Cu_3(MoO_4)_2(OH)_2$ and $Cu_3Mo_2O_9$ at room temperature. Also, the results of the magnetic properties investigation of the $Cu_3(MoO_4)_2(OH)_2$ and $Cu_3Mo_2O_9$ nanosheet samples showed that these nanosheet samples have super paramagnetic behavior at room temperature. Nanosheet morphology possess a unique feature of two-dimensional anisotropy and due to trigger quantum confinement effects hence gaining new physiochemical properties. This morphology has important applications in the field of energy conversion, storage devices, biological sensors, electronics, ferromagnetic, magneto-optical, electrochemical, photo-responsive, flexible electrochromic devices, biotechnology, water splitting, catalysis, gas sensing, and energy storage due to their unique electric properties and high packing densities [11-15].

**Experimental Section**

**Materials and Apparatus**

All of the raw materials used in this research except $Cu(NO_3)_2.3H_2O$ (Panreac) were purchased from Merck company. The structure and phase purity of as-synthesized samples was characterized by powder X-ray diffraction (XRD) on a Philips X'pert X-ray diffractometer using Cu Kα radiation (wavelength, λ = 1.5418 Å). Fourier transform infrared (FT-IR) spectra of samples were recorded on a Shimadzu-8400S spectrometer in the range of 400–4000 cm$^{-1}$ using KBr pellets. The scanning electron microscopy (SEM) analysis of lindgrenite ($Cu_3(MoO_4)_2(OH)_2$) and $Cu_3Mo_2O_9$ were recorded by TESCAN (VEGA3) instrument. Thermogravimetric analysis (TGA) and differential thermal analysis (DTA) were performed on a STA 504, the atmosphere was air. Photoluminescence (PL) spectra of the nanosheet samples were recorded on a JASCO FP-6500 spectrofluorometer at room temperature. Finally, the magnetic properties of the nanosheet samples were measured at room temperature using a vibrating sample magnetometer (VSM; Meghnatis Kavir Kashan Co., Kashan, Iran) in a maximum applied field of 15 k Oe.

**Synthesis of $Cu_3(MoO_4)_2(OH)_2$ Nanorods**

For the synthesis of lindgrenite nanorods, $Cu(NO_3)_2.3H_2O$ (0.42 g, 1.74 mmol), $Na_2MoO_4.2H_2O$ (0.28 g, 1.16 mmol) and 6-aminohexanoic acid (0.46 g, 3.51 mmol) were dissolved in 140 mL distilled water. The mixture was stirred for 10 min at room temperature. Precipitates were produced immediately from the combination of the metal cation ($Cu^{2+}$) and molybdate anion ($MoO_4^{2-}$) in pH of 5, the initial pH of solution. The resulting reaction mixture (140 mL) was transferred to 200 mL Teflon-lined stainless steel autoclave and treated to 180 ºC for 20 h. The obtained sample were filtered out and washed with distilled water, and dried at 60 °C for 2 h or more.

**Synthesis of $Cu_3(MoO_4)_2(OH)_2$ Nanosheets**

Moreover, for the synthesis of lindgrenite nanosheets, $Cu(NO_3)_2.3H_2O$ (0.23 g, 0.96 mmol) and 6-aminohexanoic acid (0.50 g, 3.84 mmol) were dissolved in 63 mL distilled water. A 0.03 M aqueous solution of sodium molybdate (63 mL) was added to the above solution at room temperature under stirring for 10 min. The resulting reaction mixture (126 mL) was transferred to 220 mL Teflon-lined stainless steel autoclave and treated to 180 ºC for 20 h. The obtained sample were filtered out and washed with distilled water, and dried at 60 °C for 2 h or more.

**Synthesis of $Cu_3(MoO_4)_2(OH)_2$ Hollow Spheres**

Finally, for the synthesis of lindgrenite hollow spheres, $Cu(NO_3)_2\cdot3H_2O$ (0.42 g, 1.74 mmol) and 6-aminohexanoic acid (0.46 g, 3.51 mmol) were dissolved in 57 mL distilled water. A 0.015 M aqueous solution of sodium molybdate (57 mL) was added to the above solution at room temperature under stirring for 10 min. The resulting reaction mixture (114 mL) was transferred to 200 mL Teflon-lined stainless steel autoclave and treated to 60 ºC for 20 h. The obtained sample were filtered out and washed with distilled water, and dried at 60 °C for 2 h or more.

**Synthesis of $Cu_3Mo_2O_9$ with Various Architectures**

For the synthesis of $Cu_3Mo_2O_9$ with various architectures, the as-synthesized $Cu_3(MoO_4)_2(OH)_2$ samples directly calcined in a furnace at calcination temperature of 500 °C, in air for 4 h with heating rate of 10 °C min$^{-1}$.

**Results and Discussion**

**Structure and Morphology of $Cu_3(MoO_4)_2(OH)_2$**

The structure of lindgrenite is composed of the of $Cu_3(OH)_2$ ribbons of the brucite structure consisting of edge-sharing copper octahedra with two kinds of copper atoms, Cu(1) and Cu(2), both having distorted octahedral coordination of oxygen atoms, running parallel to the *c* axis. $MoO_4$ connects three ribbons together via its oxygen atoms: O(1) and O(3) are bonded to single Cu atoms, while O(2) and O(4) bridges two copper atoms each on two adjacent chains [16]. Fig. 1 shows the XRD patterns of lindgrenite samples prepared by hydrothermal surfactant-assisted synthesis. Lindgrenite ($Cu_3(MoO_4)_2(OH)_2$) crystallizes in a monoclinic system with the space group of *P*2$_1$/n (No. 14) [16]. Diffraction patterns of all $Cu_3(MoO_4)_2(OH)_2$ samples are well-indexed to the monoclinic phase structure, and they are in accordance with the JCPDS card no. 75-1438 (Fig. 1). The intense diffractions in patterns without any other peaks indicate the high purity and well crystallinity of samples. The eight primary reflections; 12.62°, 20.42°, 21.34°, 24.85°, 25.34°, 25.42°, 25.62° and 33.43° are indexed as (020), (021), (101), (121), (130), (040), (111) and (141) of $Cu_3(MoO_4)_2(OH)_2$, respectively, and they are in agreement with the data reported in literature [17].

Fig. 1

The FTIR spectra of the colloidal $Cu_3(MoO_4)_2(OH)_2$ samples are shown in Fig. 2a-c. The bands at about 800-1000 $cm^{-1}$ are ascribed to $MoO_4^{2-}$ stretching vibration, and it is consistent with the reported values [9]. The band at about 450 $cm^{-1}$ is assigned to the bending vibration of Cu-O. Also, the weak absorptions at about 1600 $cm^{-1}$ and the bands at 3000 –3600 $cm^{-1}$ demonstrate the bending and the stretching vibrations of $H_2O$ and O-H, respectively [18-19]. Bands at about 2800-2900 $cm^{-1}$ attributed to the C-H stretching vibrations of methylene groups of the AHA molecules which have been overlapped with the stretching vibration of $H_2O$ bands, and bands at about 1400 $cm^{-1}$ correspond to C-N bending modes of AHA molecules. Meanwhile, Bands at about 3400 $cm^{-1}$ correspond to $-NH_2$ stretching mode of AHA molecules [20-23]. These results demonstrate that the carboxylic (-COOH) terminus are free and thus, approve hydrophilicity of the three samples.

Fig. 2

The morphologies of as-prepared products were also examined by SEM (Fig. 3). Fig. 3a-b show SEM images of the prepared $Cu_3(MoO_4)_2(OH)$, having rod morphology with diameters of about 100 nm. Fig. 3c-e show SEM image of the $Cu_3(MoO_4)_2(OH)_2$ prepared with a 0.03 M aqueous solution of sodium molybdate, at 180 °C. As Figures show the as-synthesized sample have sheet forms with the sizes of 30−100 nm in width. The SEM observation also indicates that almost 100% of the obtained $Cu_3(MoO_4)_2(OH)_2$ sample are super thin sheet and no other type of particles is visible. Fig. 3f shows SEM image of the $Cu_3(MoO_4)_2(OH)_2$ prepared with a 0.015 M aqueous solution of sodium molybdate, at 60 °C in which the as-synthesized sample have hallow sphere architecture. The mean diameter of these hallow spheres were about 25 $\mu$m as shown in Fig. 3g. Moreover, SEM image of the $Cu_3(MoO_4)_2(OH)_2$ hallow spheres demonstrate which an individual hallow sphere is composed of a large number of nanosheets with a thickness of about 40-70 nm (Fig. 3h). Also, the SEM observation shows that the spheres have hollow structures (Fig. 3i)

Fig. 3

## Possible Growth Mechanism of $Cu_3(MoO_4)_2(OH)_2$ with Various Architectures

For $Cu_3(MoO_4)_2(OH)_2$ nanorods, purposed mechanism based on literature reports, involves two mechanisms well known as Ostwald ripening and oriented attachment [24-25]. When $Cu(NO_3)_2.3H_2O$ solution was added to the $MoO_4^{2-}$ solution in the presence of 6-aminohexanoic acid, $Cu(OH)_2$ formed firstly. Then $Cu(OH)_2$ combined with copper ions and $MoO_4^{2-}$ and a high supersaturation degree will be reached, and amorphous $Cu_3(MoO_4)_2(OH)_2$ particles will form immediately. This process is described in Equations (1) and (2) [26].

$Cu^{2+} + 2OH^- \rightarrow Cu(OH)_2$ (1)

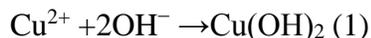

$Cu(OH)_2 + 2Cu^{2+} + 2MoO_4^{2-} \rightarrow Cu_3(MoO_4)_2(OH)_2$ (2)

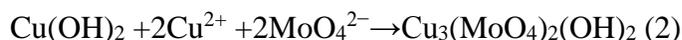

When the synthesis temperature was increased, monodisperse spherical nanoparticles were obtained. At a longer reaction time, the equilibrium of growth kinetics was observed and at higher reaction temperature, according to the well-known Ostwald ripening process, a mixture of spherical particles and primary rod-like particles were observed and finally, with the temperature increasing of the reaction no nanoparticles were observed. At elevated temperature, the nanorods involved thicker and longer colloidal nanorods. At this stage, the oriented attachment is dominant mechanism [24].

On the other hand, for $Cu_3(MoO_4)_2(OH)_2$ nanosheets, based on literature reports [27-28], was considered that the molybdate nanosheets formation can be the effect of among Ostwald ripening [29], lateral-aggregation [24], and dissolution/recrystallization [27] processes. Probably, this is due to the presence of 6-aminohexnoic acid molecules in the water medium. Similar to rod morphology, when $Cu(NO_3)_2.3H_2O$ solution was added to the $MoO_4^{2-}$ solution in the presence of 6-aminohexanoic acid, $Cu(OH)_2$ formed firstly. Then $Cu(OH)_2$ combined with copper ions and $MoO_4^{2-}$ and a high supersaturation degree will be reached, and amorphous $Cu_3(MoO_4)_2(OH)_2$ particles will form immediately. Apparently, the interesting transformation of a small crystalline nucleus into lindgrenite nanoparticles in a supersaturated solution at elevated temperature, proceed through the process known as Ostwald ripening. Subsequently, the primary nanoparticles quickly transform into lindgrenite nanosheets through the lateral-aggregation mechanism. At the end of the synthetic process, the shape transformation of the crystallized nanosheets often operates through a standard dissolution-recrystallization mechanism. Thus, possible growth mechanism of the $Cu_3(MoO_4)_2(OH)_2$ nanosheets was dominated by crystallization-dissolution-recrystallization mechanism.

For self-assembled $Cu_3(MoO_4)_2(OH)_2$ hollow sphere sample, after formation of the subunit nanosheets *via* the purposed mechanism for lindgrenite nanosheets, the assembly mechanism of sphere-like architecture operate. Therefore, possible growth mechanism of the $Cu_3(MoO_4)_2(OH)_2$ hollow sphere morphology was dominated by crystallization-dissolution-recrystallization- self-assembly growth mechanism.

**Structure and Morphology of $Cu_3Mo_2O_9$**

Tertiary copper molybdate, $Cu_3Mo_2O_9$, have an orthorhombic structure with space group *Pnma*. In the unit cell of this material two types of Mo–4O tetrahedra, one compressed Cu-6O octahedral, and two types of Cu-5O polyhedral exist. The catalytic properties of this metal oxide are closely related to metal-oxygen bonds [2].

XRD patterns of the obtained $Cu_3Mo_2O_9$ with various architectures at calcination temperature of 500 °C are shown in Fig. 4a-c. Diffraction patterns of the three $Cu_3Mo_2O_9$ samples are well-indexed to the orthorhombic structure, and they are in accordance with the standard diffraction patterns of $Cu_3Mo_2O_9$, the JCPDS card no. 70-2493.

Fig. 4

The FTIR spectra of the colloidal $Cu_3Mo_2O_9$ samples are shown in Fig. 5a-c. Due to the difference of Cu–O and Mo–O bands in the crystal structures of $Cu_3(MoO_4)_2(OH)_2$ and $Cu_3Mo_2O_9$, in the spectra of $Cu_3Mo_2O_9$ samples, the excess bands were appeared in the range of 400–1300 cm$^{-1}$ [30-32]. The bands at 530 cm$^{-1}$ is assigned to the bending vibration of Cu−O. Also, the absorptions at 1600 cm$^{-1}$ and the bands at 3000 –3700 cm$^{-1}$ demonstrate the bending and the stretching vibrations of $H_2O$ and O-H, respectively [18-19]. Bands at about 2800-2900

cm$^{-1}$ attributed to the C-H stretching vibrations of methylene groups of the AHA molecules and band at about 1400 cm$^{-1}$ correspond to C-N bending modes of AHA molecules, respectively. Bands at about 3400 cm$^{-1}$ correspond to -NH$_2$ stretching mode of AHA molecules which have been overlapped with the stretching vibration of O-H bands [20-23]. These results approve hydrophilicity of the three samples. As seen in Fig. 2, comparing with the bands of O-H stretching modes in Cu$_3$Mo$_2$O$_9$ samples, due to the existence of hydrogen bond in the lindgrenite crystal lattice, the bands of O-H stretching modes in Cu$_3$(MoO$_4$)$_2$(OH)$_2$ samples have shifted to the lower wavenumbers and split into two bands.

Fig. 5

SEM images of the Cu$_3$Mo$_2$O$_9$ samples obtained at calcination temperature of 500 °C, are shown in Fig. 6. SEM image of the Cu$_3$Mo$_2$O$_9$ obtained from calcined Cu$_3$(MoO$_4$)$_2$(OH)$_2$ nanorod sample is shown in Fig. 6a. As Figure shows the as-synthesized sample have rod forms with diameters of about 100 nm. Also, SEM image of the Cu$_3$Mo$_2$O$_9$ obtained from calcined Cu$_3$(MoO$_4$)$_2$(OH)$_2$ nanosheet sample is shown in Fig. 6b-c. This sample has retained the morphology of the lindgrenite nanosheet sample with the sizes of 20−100 nm in thickness. Finally, for Cu$_3$Mo$_2$O$_9$ hollow sphere sample, calcined at 500 °C, the morphology of the lindgrenite hollow sphere sample has been retained with the same size in diameter of hollow sphere and nanosheet thickness, contributing in hollow sphere structure, as shown in Fig. 6d-e.

Fig. 6

**Thermal Stability of $Cu_3(MoO_4)_2(OH)_2$ and $Cu_3Mo_2O_9$ Nanosheets**

In order to investigation of the thermal stability of $Cu_3(MoO_4)_2(OH)_2$ and $Cu_3Mo_2O_9$, TG-DTA curves obtained under a flow of air as shown in Fig. 7. According to thermogravimetry (TG) analysis, the weight loss consists of two distinct steps. In the first stage, the TG curve indicated a weight loss (about 3.41 Wt%) between 50 °C and 450 °C, corresponding to the loss of adsorbed water and coordinated water [33] and elimination of OH groups as $H_2O$, result in the formation of $Cu_3Mo_2O_9$. Meanwhile, weight-loss step in the range of 135 to 450 °C, corresponding to combustion of 6-aminohexanoic acid of $Cu_3(MoO_4)_2(OH)_2$ surface [34]. A second weight loss observed between 700 °C and 800 °C, corresponding to the decomposition of $Cu_3Mo_2O_9$. The three endothermic peaks in DTA curve at 171 °C, 384 °C and 387 °C are assigned to the loss of adsorbed water, coordinated water and 6-aminohexanoic acid of $Cu_3(MoO_4)_2(OH)_2$, respectively. Finally, significant endothermic peak at 780 °C resulted from decomposition of $Cu_3Mo_2O_9$.

Fig. 7

**Photoluminescence Properties of $Cu_3(MoO_4)_2(OH)_2$ Nanosheets**

According to the earlier reports, $Cu_3(MoO_4)_2(OH)_2$ has the monoclinic structure with space group P$2_1/n$ and its crystal structure consists of strips of edge-sharing $CuO_6$ octahedral that are cross-linked by $MoO_4^{2-}$ tetrahedral [3]. The $MoO_4^{2-}$ tetrahedral is represented as $\Gamma_{Td}= A_1(v_1) + E(v_2) + F_2(v_3) + F_2(v_4)$ in which $A_1(v_1)$ and $E(v_2)$ are Raman active, but $F_2(v_3)$ and $F_2(v_4)$ are infrared active. Since the $F_2(v_3)$ vibrations are antisymmetric stretches, the bands at about 800 cm$^{-1}$ are assigned to $F_2(v_3)$ antisymmetric stretching vibrations of $Cu_3(MoO_4)_2(OH)_2$. In this study, all samples exhibit the same bands at about 813 cm$^{-1}$. This band could be originated from the Mo-O stretching vibration of the $MoO_4$ group [35].

Fig. 8 shows the room-temperature photoluminescence spectrum of the as-prepared $Cu_3(MoO_4)_2(OH)_2$ nanosheetss using the excitation wavelength of 237 nm. The spectrum shows that Lindgrenite nanosheets exhibited emission peaks at 705 nm which could be due to the charge-transfer transitions within the $MoO_4^{2-}$ complex [36-40]. Compared with earlier results, $Cu_3(MoO_4)_2(OH)_2$ sample that synthesized in this work by the surfactant-assisted hydrothermal method has a very obvious red-shifted PL emission. It is well known that the differences of maximum emission wavelengths can be assigned to the structural organization levels, preparation methods, treatment conditions, and different excitation wavelengths and the relative intensity of the emission peaks closely related to the morphology, size, surface defect states, and so on [41-42]. Generally, these results indicate that high crystalline hydrophilic surfaced $Cu_3(MoO_4)_2(OH)_2$ nanosheets prepared in this work have potential as a photoluminusent material.

Fig. 8

**Magnetic Properties of $Cu_3(MoO_4)_2(OH)_2$ Nanosheets**

The magnetization of our $Cu_3(MoO_4)_2(OH)_2$ nanosheets sample obtained by surfactant-assisted hydrothermal method as a function of magnetic field at the maximum field of 10 kOe is shown in Fig. 9. The magnetization of this sample is not completely saturated yet at the maximum field of the measurements (10 kOe). Hysteresis loops at room temperature are much thin and reveal that the as-synthesized $Cu_3(MoO_4)_2(OH)_2$ nanosheets exhibits zero coercivity and, thus, have superparamagnetic at room temperature.

Fig. 9

**Photoluminescence Properties of Cu$_3$Mo$_2$O$_9$ Nanosheets**

According to the earlier reports, Cu$_3$Mo$_2$O$_9$ has an orthorhombic structure with space group *Pnma* and in the unit cell of this material exists two types of Mo-4O tetrahedral [2]. Fig. 10 shows the room-temperature photoluminescence spectrum of the as-prepared Cu$_3$Mo$_2$O$_9$ nanosheets using the excitation wavelength of 237 nm. The spectrum shows that nanosheets exhibited emission peaks at 706 nm which could be due to the charge-transfer transitions within the MoO$_4^{2-}$ complex [36-40]. The as-synthesized Cu$_3$Mo$_2$O$_9$ sample by the surfactant-assisted hydrothermal method also has a very obvious red-shifted PL emission. Generally, these results indicate that Cu$_3$Mo$_2$O$_9$ nanosheets prepared in this work similar to high crystalline hydrophilic surfaced Cu$_3$(MoO$_4$)$_2$(OH)$_2$ nanosheets have potential as a photoluminusent material.

Fig. 10

**Magnetic Properties of Cu$_3$Mo$_2$O$_9$ Nanosheets**

The magnetization of as-prepared Cu$_3$Mo$_2$O$_9$ nanosheet sample at calcination temperature of 500 °C as a function of magnetic field is shown in Fig. 11. The magnetization of this sample is not completely saturated yet at the maximum field of the measurements (15 kOe). Hysteresis loops at room temperature are much thin and reveal that the as-synthesized sample exhibits zero coercivity and, thus, have superparamagnetic at room temperature. Moreover, the saturation magnetization $M_s$ of as-synthesized Cu$_3$Mo$_2$O$_9$ is higher than as-synthesized Cu$_3$(MoO$_4$)$_2$(OH)$_2$ at room temperature.

Fig. 11

**Conclusion**

In this study, well-crystalline monoclinic $Cu_3(MoO_4)_2(OH)_2$ and orthorhombic $Cu_3Mo_2O_9$ nanostructures have been prepared *via* a simple and green route. This method is highly scale up able. The results demonstrated that well crystallized lindgrenite samples have rod morphology with diameters of about 100 nm, sheet form with thickness of 30–100 nm and hallow sphere shape, consisting of a large number of nanosheets with thickness of about 40-70 nm. The fabricated $Cu_3Mo_2O_9$ samples from thermal treatment of lindgrenite retained the original morphologies. Furthermore, possible mechanisms for the formation of hydrophilic-surfaced $Cu_3(MoO_4)_2(OH)_2$ crystals is proposed. Possible proposed mechanisms of the $Cu_3(MoO_4)_2(OH)_2$ with various architectures, were dominated by Ostwald ripening-oriented attachment, crystallization-dissolution-recrystallization and crystallization-dissolution-recrystallization-self-assembly growth mechanisms for nanorod, nanosheet and hollow sphere morphologies, respectively. Meanwhile, the photoluminescence and magnetic properties of the nanosheet samples have been investigated that the both of $Cu_3(MoO_4)_2(OH)_2$ and $Cu_3Mo_2O_9$ samples have super paramagnetic behavior at room temperature and in comparison with previous works, $Cu_3(MoO_4)_2(OH)_2$ and $Cu_3Mo_2O_9$ samples synthesized by the surfactant-assisted hydrothermal method in this work have a very obvious red-shifted PL emission and high intensity.

**Acknowledgment.** The authors acknowledge finical support of Tarbiat Modares University and Shahid Beheshti University for supporting this work.

**Fig. Captions**

**Fig. 1.** XRD patterns for the $Cu_3(MoO_4)_2(OH)_2$ samples with (a) rod, (b) sheet and (c) hollow spherical shapes.

**Fig. 2.** Infrared spectra of the as-prepared colloidal $Cu_3(MoO_4)_2(OH)_2$ with various architectures: (a) rod, (b) sheet and (c) hollow sphere morphology.

**Fig. 3.** SEM images of (a-b) $Cu_3(MoO_4)_2(OH)_2$ rod-shaped structure, (c-e) $Cu_3(MoO_4)_2(OH)_2$ sheet form structure, (f-i) $Cu_3(MoO_4)_2(OH)_2$ hollow spherical shape.

**Fig. 4.** XRD patterns of the $Cu_3Mo_2O_9$ samples calcined at 500 °C with (a) rod, (b) sheet and (c) hollow sphere morphologies.

**Fig. 5.** Infrared spectra of the colloidal $Cu_3Mo_2O_9$ nanostructures: (a) rod, (b) sheet and (c) hollow sphere morphology.

**Fig. 6.** SEM images of the colloidal $Cu_3Mo_2O_9$ (a) nanorods, (b-c) nanosheets and (d-e) hollow spheres.

**Fig. 7.** TAG and DTA analysis of the prepared $Cu_3(MoO_4)_2(OH)_2$ and $Cu_3Mo_2O_9$ nanosheets.

**Fig. 8.** Room-temperature photoluminescence spectrum of the prepared $Cu_3(MoO_4)_2(OH)_2$ nanosheet sample.

**Fig. 9.** Magnetic properties of the lindgrenite nanosheets: magnetization vs. field and hysteresis loops of sample at room temperature.

**Fig. 10.** Room-temperature photoluminescence spectrum of the $Cu_3Mo_2O_9$ nanosheets (calcined at 500 °C).

**Fig. 11.** Magnetic properties of the $Cu_3Mo_2O_9$ nanosheets: magnetization vs. field and hysteresis loops of sample at room temperature.

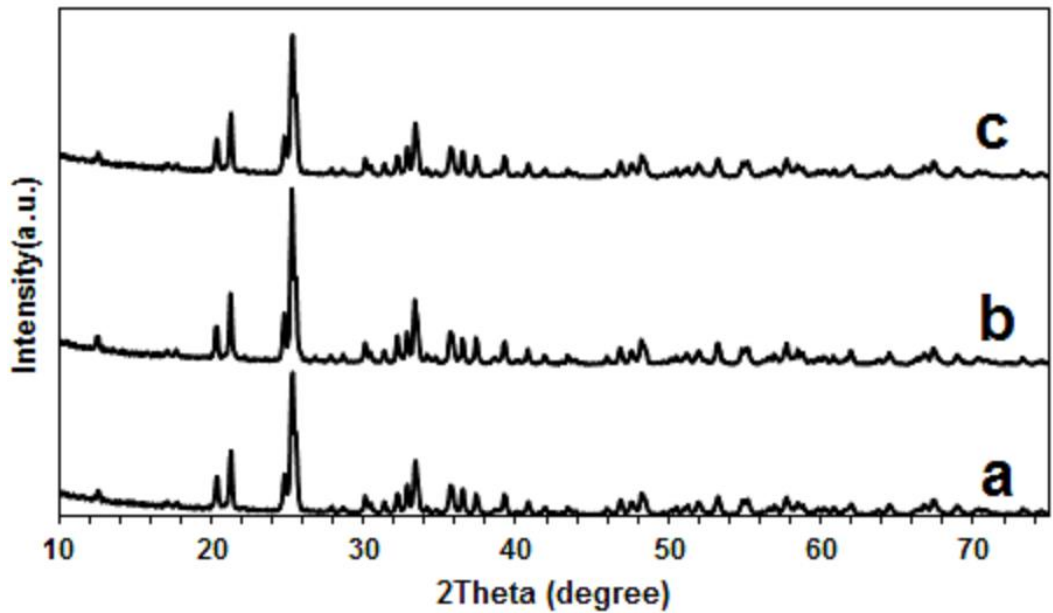

**Fig. 1**

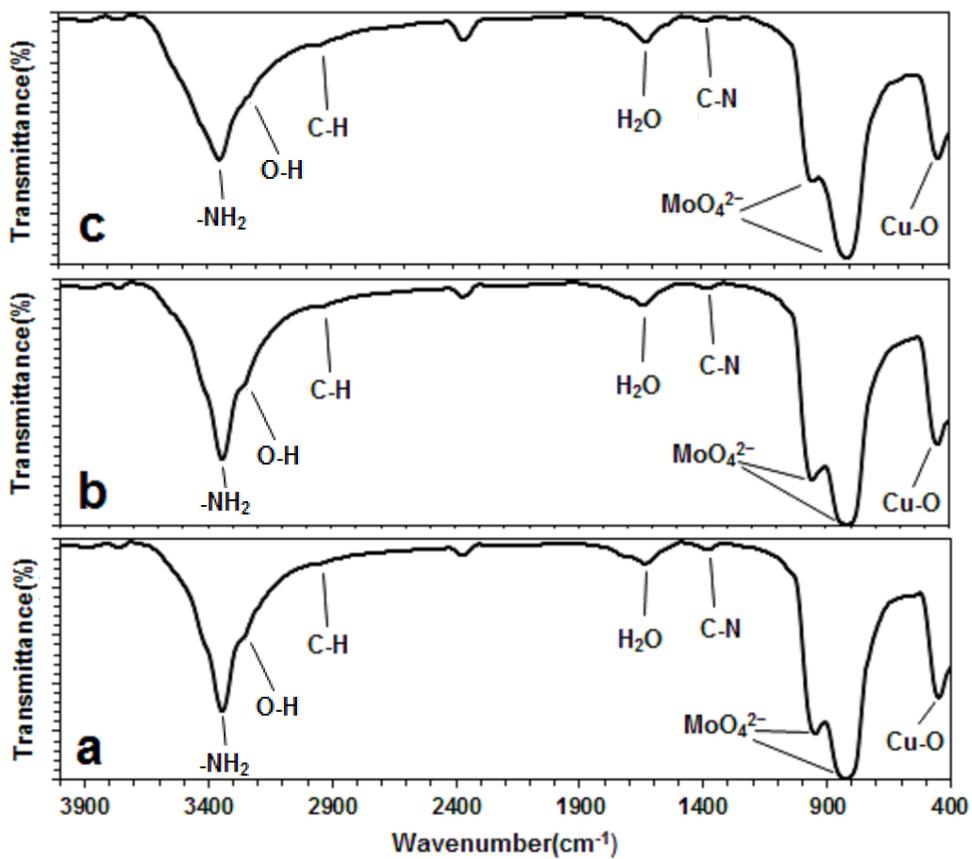

**Fig. 2**

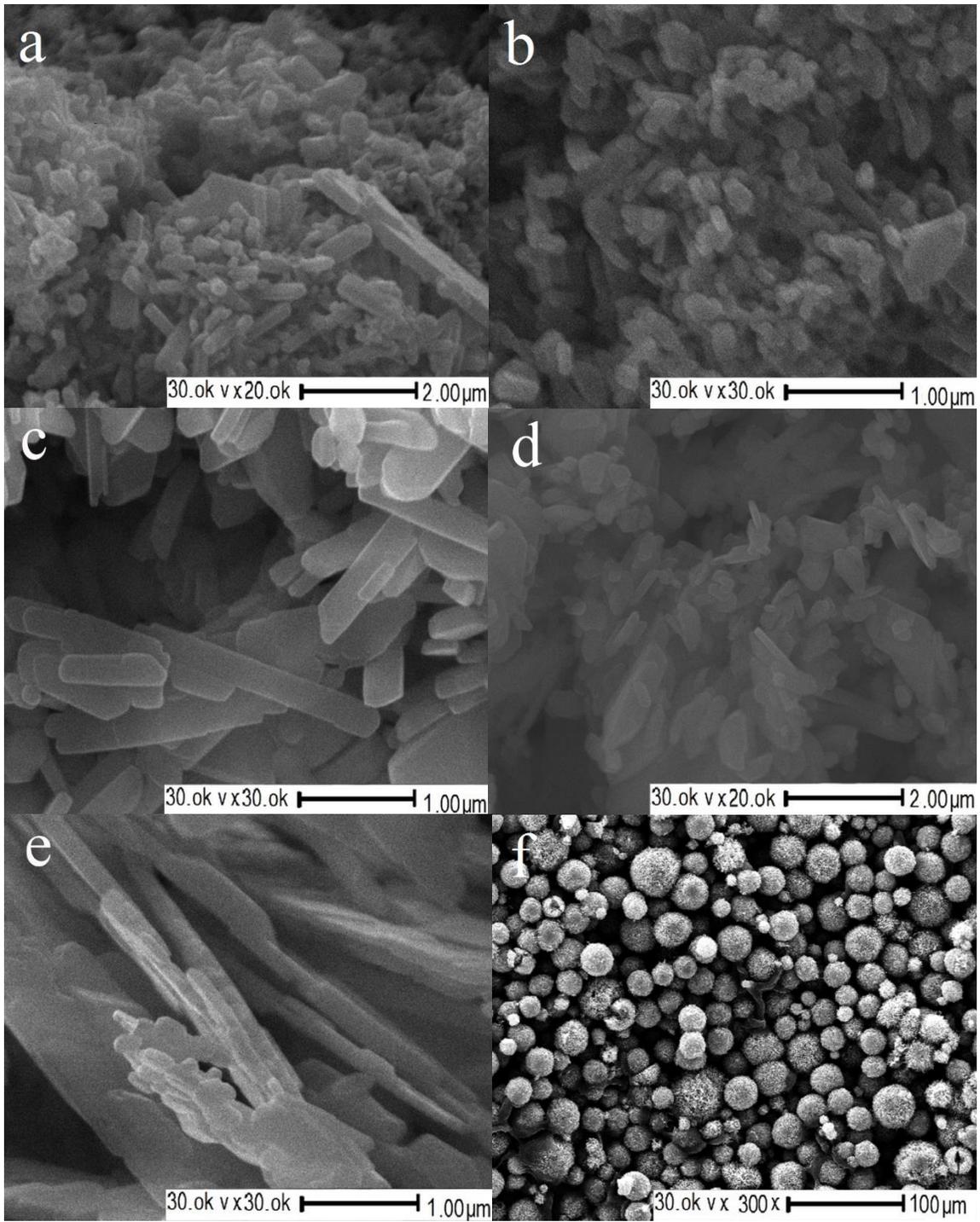

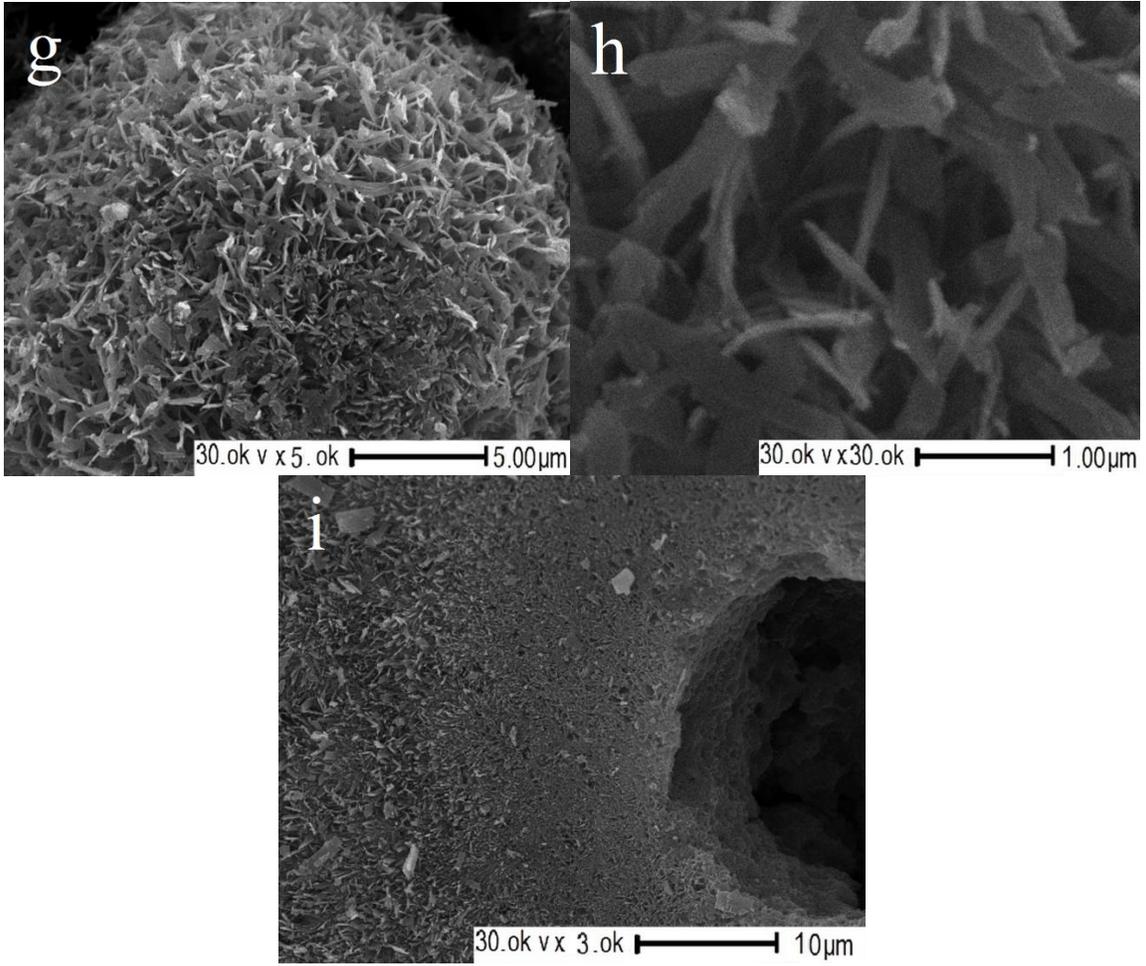

**Fig. 3**

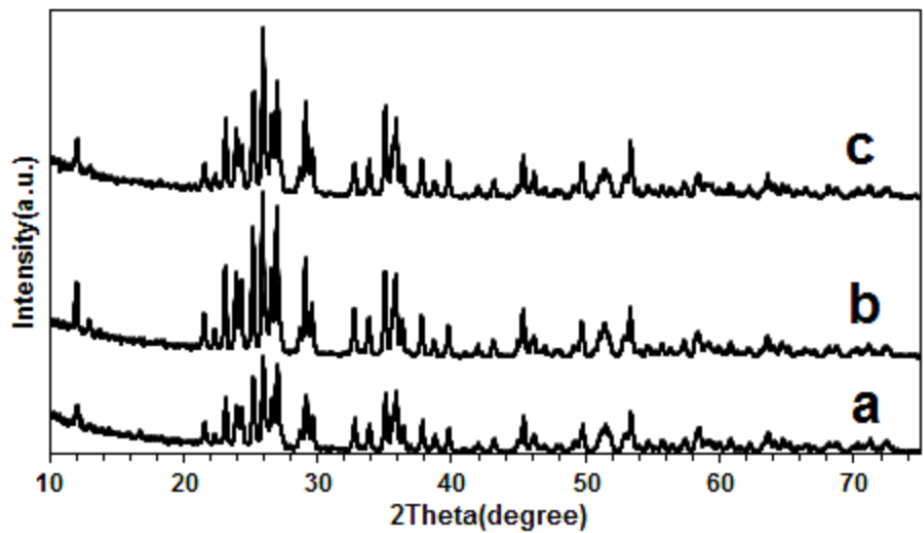

**Fig. 4**

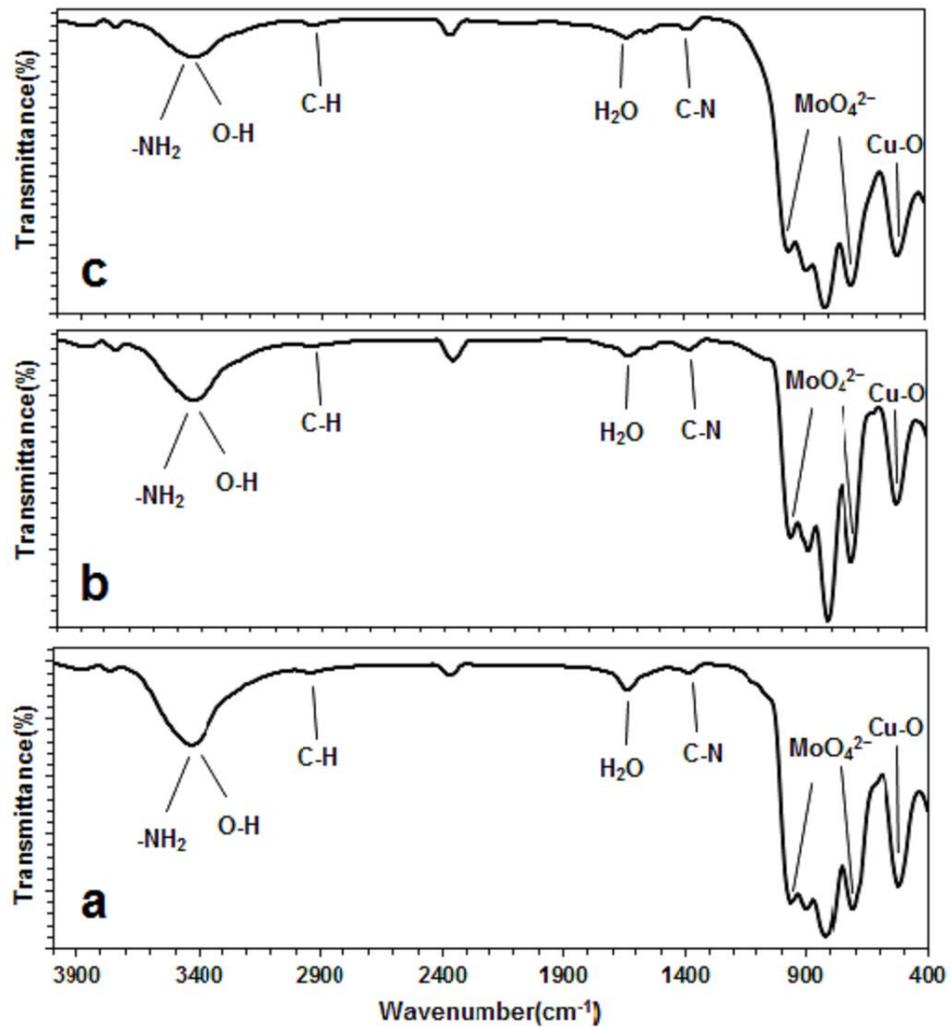

**Fig. 5**

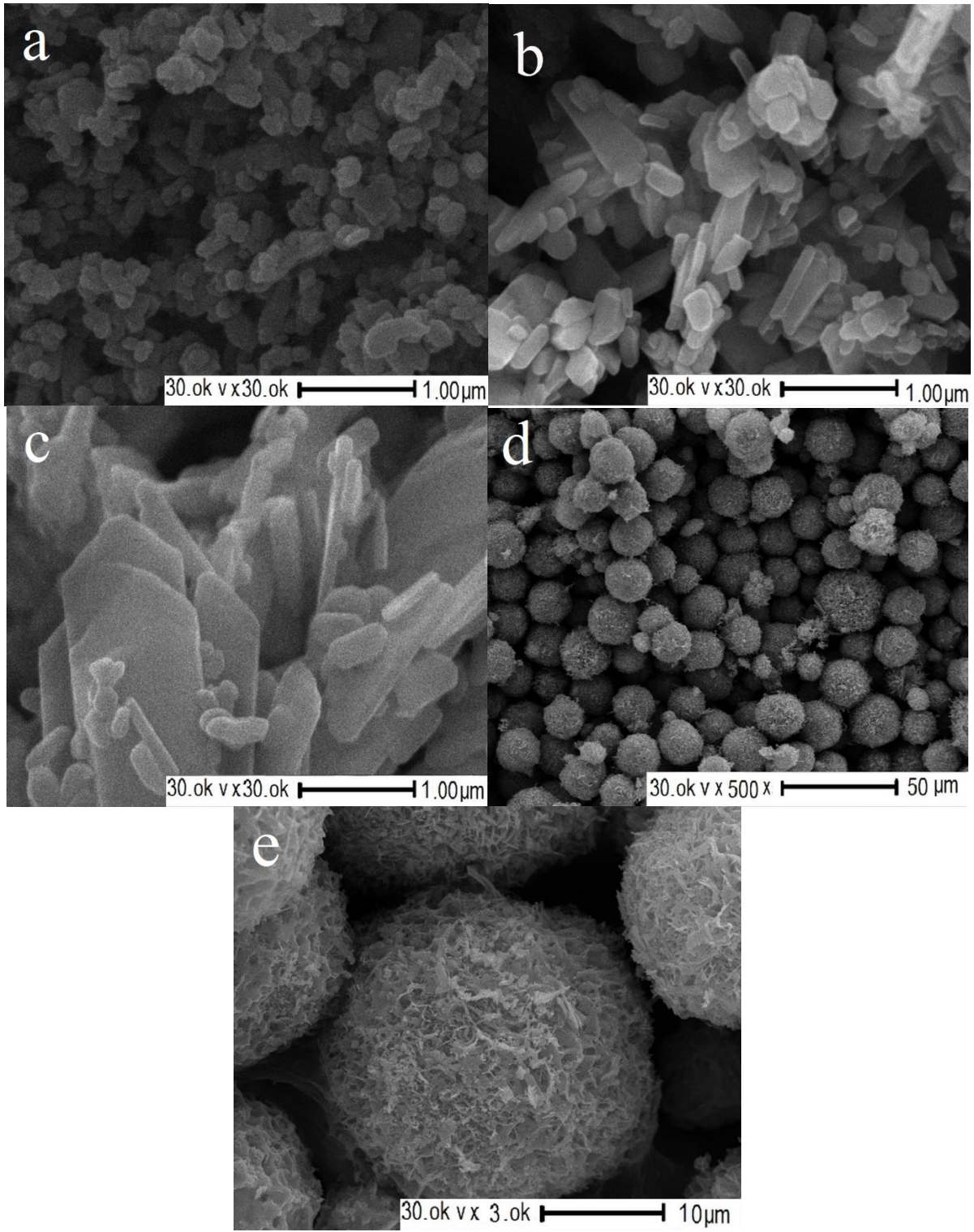

**Fig. 6**

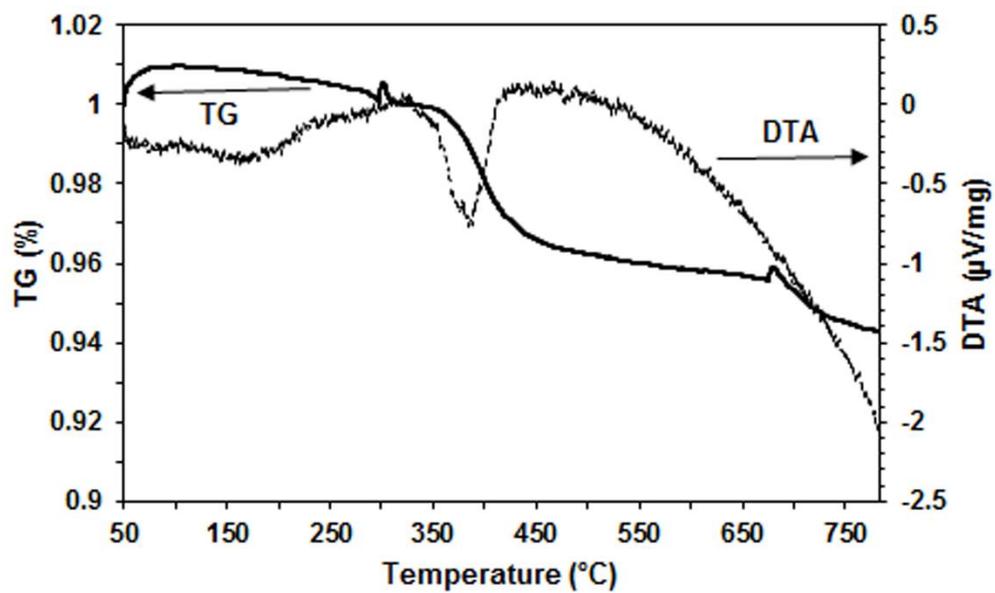

**Fig. 7**

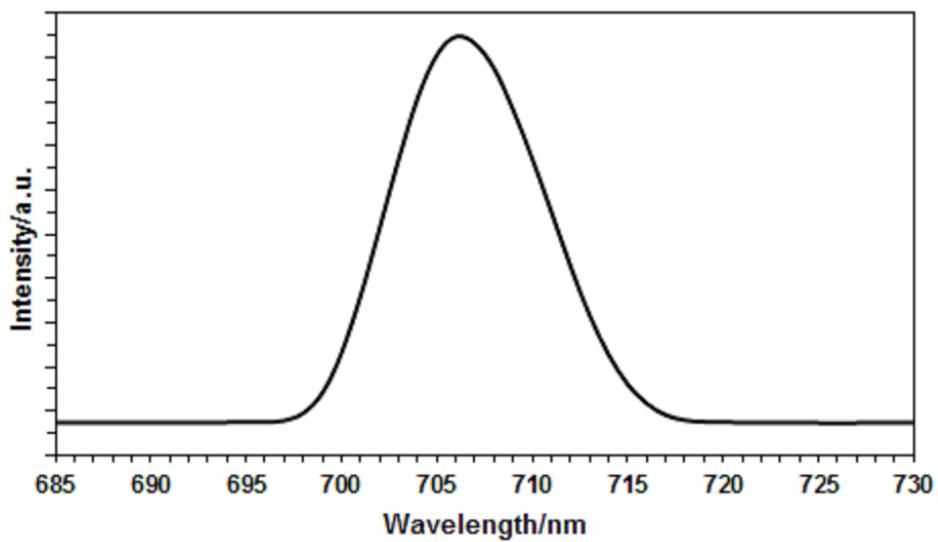

**Fig. 8**

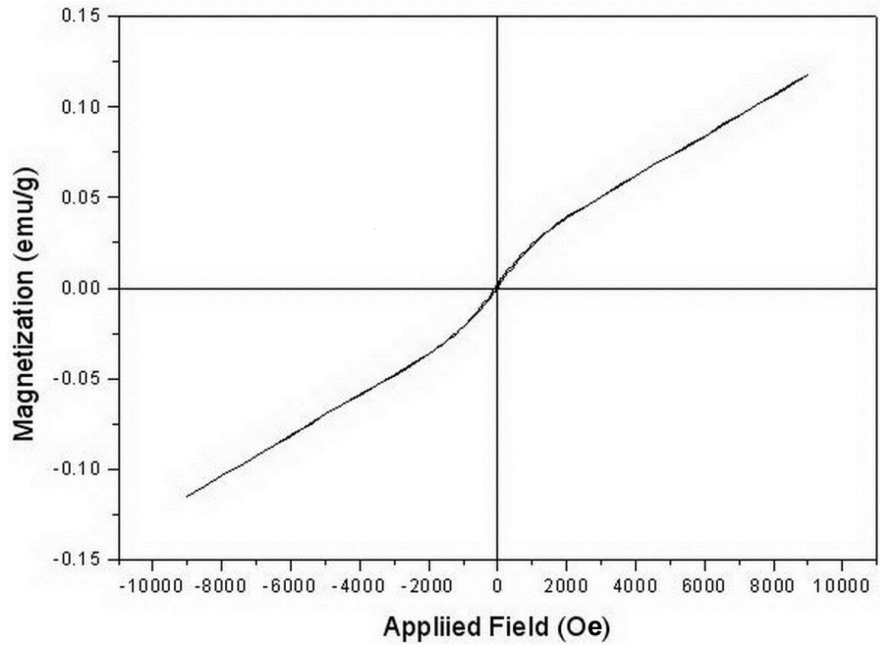

**Fig. 9**

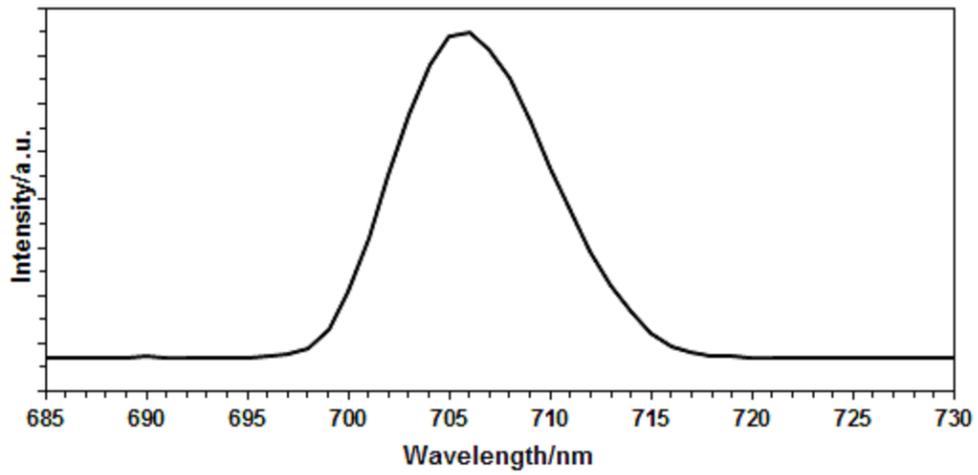

**Fig. 10**

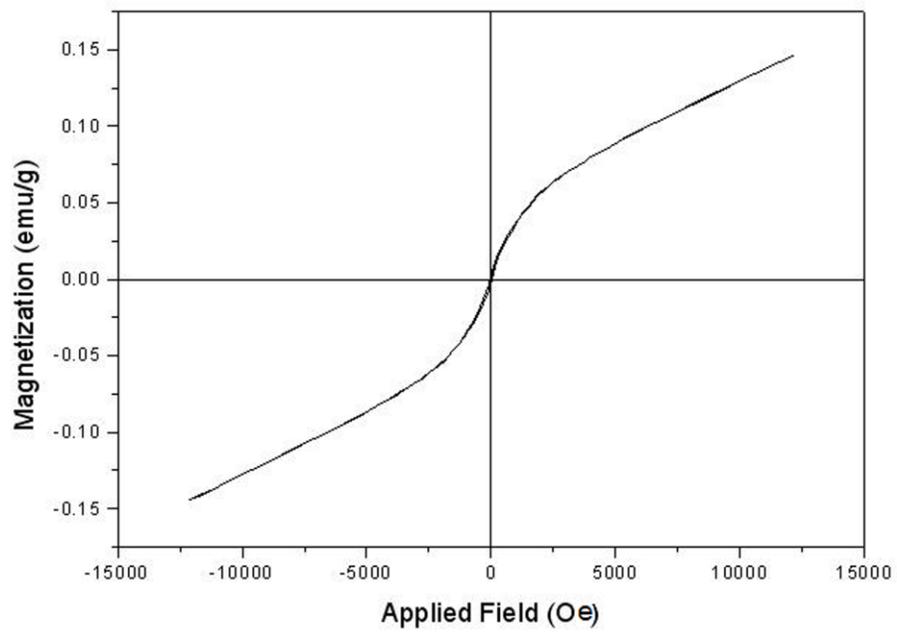

**Fig. 11**